\documentclass{appolb}
\usepackage{epsfig}

\newcommand{\kp}{\mbox{$k_{\perp}$}}

\newcommand{\kpijq}{\mbox{$k^{2}_{\perp ij}$}}
\newcommand{\ycut}{\mbox{$y_{\rm cut}$}}
\newcommand{\yt}{\mbox{$y_2$}}
\newcommand{\QQ}{\mbox{$Q^{2}$}}
\newcommand{\alpsmz}{\mbox{$\alpha_s(M_Z)$ }}
\newcommand{\etbreit}{\mbox{$\overline{E}_{T,{\rm Breit}}$}}
\newcommand{\xp}{\mbox{$x_{p}$}}
\newcommand{\zp}{\mbox{$z_{p}$}}

\begin{document}
\pagestyle{plain}
\newcount\eLiNe\eLiNe=\inputlineno\advance\eLiNe by -1
\title{ELECTROPRODUCTION OF DIJETS \\
AT SMALL JET SEPARATION%
\thanks{Presented on behalf of the H1 collab. at DIS 2002, Krak\'ow, April 30 - May 4, 2002}%
}
\author{G\"unter Grindhammer
\address{Max-Planck-Institut f\"ur Physik, 
Werner-Heisenberg-Institut, M\"unchen
}}
\maketitle

\begin{abstract}
Dijet production in deep-inelastic scattering (DIS) in the range $150
< Q^{2} < 35000$~GeV$^{2}$ has been measured by the H1 collaboration
using the Durham jet algorithm in the laboratory frame.  QCD
calculations in next-to-leading order (NLO) are found to give a good
description of the data when requiring a small minimum jet separation,
which selects a dijet sample containing 1/3 of DIS events in contrast
to approximately 1/10 with more typical jet analyses.
\end{abstract}

\section{Introduction}

One of the remarkable results of the measurements of inclusive cross
sections in DIS at HERA is that the data can be sucessfully described
by perturbative QCD calculations down to quite small momentum
transfers squared \QQ. Investigations of the hadronic final state have
shown that QCD is also able to describe events containing two
energetic jets.  Such investigations tended to require large jet
separation, \ie large relative jet transverse momentum or large
transverse jet energy in the Breit frame.  These requirements
typically result in a sample consisting of 1/10 of DIS events.  Here
I present the results~\cite{h1-dijets} of an investigation on the
minimum jet separation needed in order to still successfully describe
the data by perturbative QCD.

\section{Modified Durham jet algorithm}

The definition of minimum jet separation used derives from the
criterion used by the Durham \kp~jet algorithm~\cite{durham-ktalgo}. 
This algorithm was invented to improve the JADE jet algorithm, widely
and successfully used in the study of $e^{+}e^{-}$ annihilation.  The
JADE algorithm uses the invariant mass between partons or particles as
a measure of jet separation.  
It was noticed that for certain
configurations 2 soft gluons are combined into a jet instead of
associating each of the gluons to the two energetic partons in the
event.  This was remedied by the Durham \kp~algorithm.  For DIS the
algorithm was further modified by adding a proton remnant pseudo
particle (missing momentum in the proton direction) in order to better
deal with initial state radiation by the colliding proton.  

The algorithm runs through the following steps:
a) compute all combinations 
   $\kpijq = 2 \min [E^{2}_{i},E^{2}_{j}](1-\cos\theta_{ij})$, 
b) combine the pair with the minimum \kpijq by adding the 
   four-momenta,
c) iterate this procedure until \eg exactly $(2+1)$~jets, \ie 2 final 
   state jets and the proton remnant jet remain, then
d) define {\em significant} dijets by imposing a lower limit \ycut~on 
   $\yt = \kpijq /\rm{scale}^2$ with the total hadronic energy $W$ or 
   $Q$ typically chosen as scale.
In the analysis presented here $W$ is chosen for the scale, which has
the advantage that the uncertainty of the hadronic energy scale of the
calorimetric energy measurement partially cancels in \yt.  The
algorithm is applied in the laboratory frame with the further
advantage of not having an additional boost error.

\section{Previous results}

Three different previous examples of measurements and applications of
jet observables as a function of \ycut are briefly discussed.  

The jet rates $R_{1+1}$, $R_{2+1}$, and $R_{3+1}$ have been presented
in ref.~\cite{zeus-alps}.  The data are well described by NLO QCD in
the measured range $0.01 \leq \ycut \leq 0.06$ and allowed a
determination of the strong coupling using $R_{2+1}$ for $\ycut =
0.02$.

The 2nd example chosen is a measurement of subjet multiplicities in
the range $0.001 \leq \ycut \leq 0.1$ presented in
ref.~\cite{h1-subjets} for jets found in the Breit frame using the
inclusive \kp~jet algorithm and demanding $E_{\rm T,jet} > 8$~GeV.
These data are well described by Monte Carlo (MC) models incorporating
leading order (LO) matrix elements, parton showers, and hadronization.

Finally, in a paper on event shapes~\cite{h1-evshapes} normalized
dijet cross sections have been shown as a function of $y$-values for
which the transition from $(2+1)$ to $(1+1)$ jets occurs.  These data
are of interest in measuring power corrections and the strong
coupling.  In most of the range $0.05 < \yt < 1$ and for not too small
\QQ, the data are well described by NLO QCD.

These examples show that perturbative QCD is able to describe
\yt~distributions down to small values corresponding to \kp of a few
GeV and that MC models can even describe subjet multiplicities down to
$\kp \approx 0.3$~GeV. This motivates the investigation presented
here: what is the minimum \yt~for which NLO QCD gives a good
description of the characteristic observables of dijets?  In addition
one may ask how well are they described by MC models.

\section{Data and dijet selection}

About $6\,10^{4}$ DIS events are selected in the range $150 < \QQ
< 35000$~GeV$^{2}$ and $0.1 < y < 0.7$ for an integrated luminosity of
35~pb$^{-1}$.  Dijet events are found in the laboratory frame as a
function of $\yt = \min \kpijq / W^{2}$ by applying the modified
Durham jet algorithm.  In addition the jets have to lie within
$10^{\circ} < \theta_{\rm jet} < 140^{\circ}$ to be well measured.

The dominant systematic errors of $\approx 5\%$, $4\%$, and $ < 1\%$
are due to the model dependence of the corrections and the uncertainty
of the hadronic and electromagnetic energy scales of the calorimeters.

\section{Results}

\begin{figure}[htb]
\epsfxsize=14pc 
\epsfbox{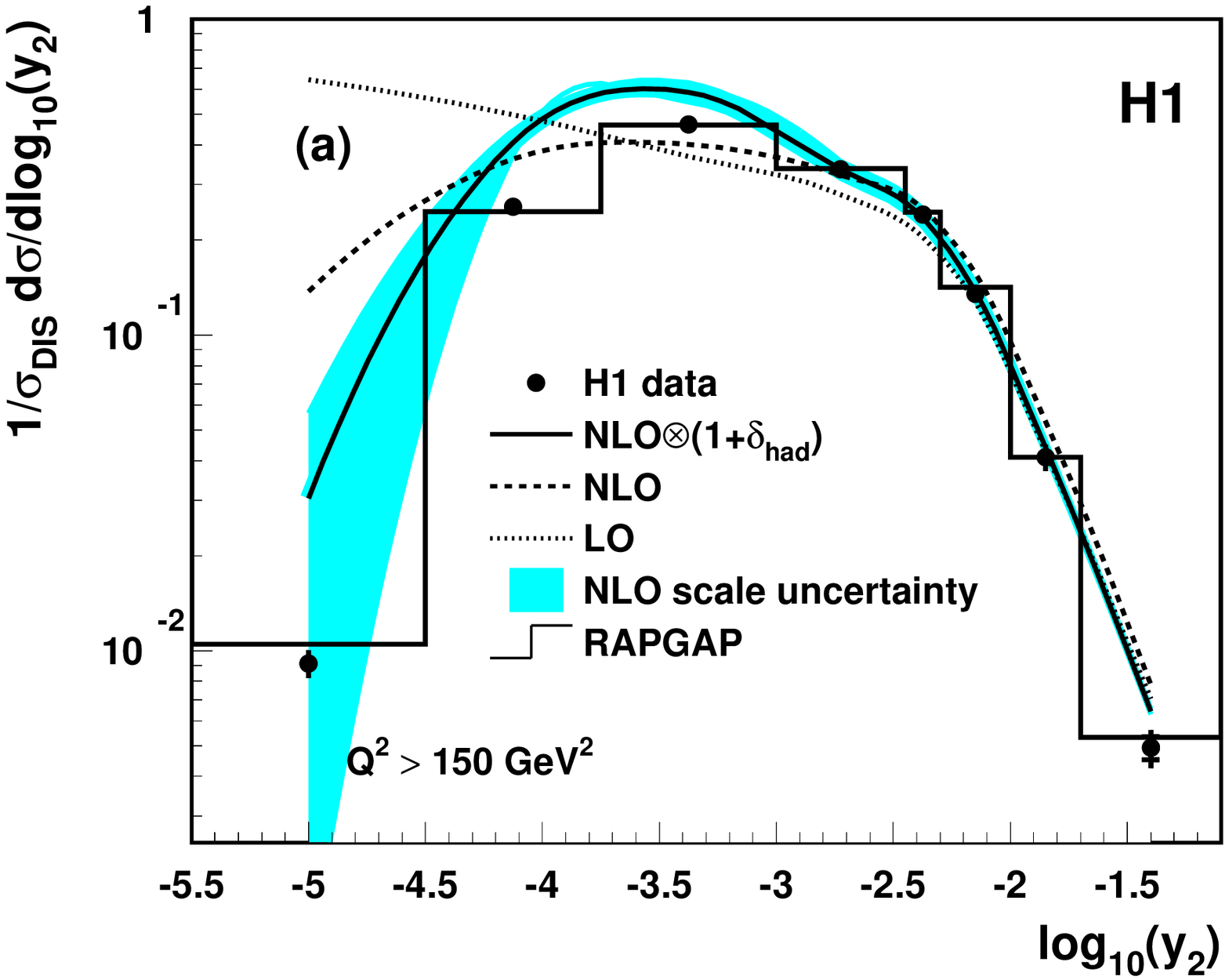}
\epsfxsize=15pc
\epsfysize=10.6pc
\epsfbox{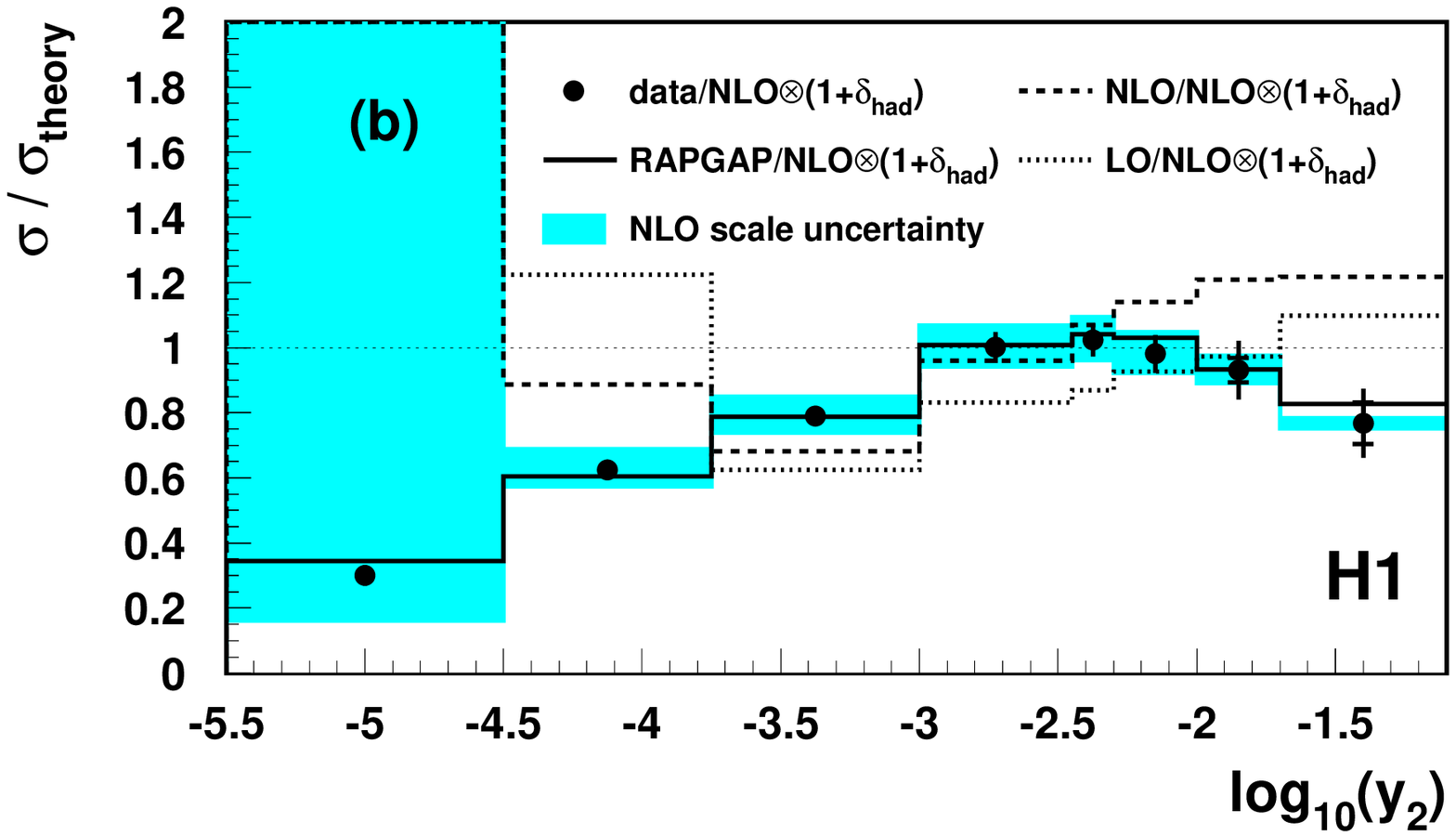}
\caption{ {\bf a} The normalized dijet cross section. Here and in the 
following figure the statistical errors are given by the inner error 
bars and the outer error bars correspond to the quadratic sum of the 
statistical and systematic errors. Also shown are perturbative QCD
predictions in LO, in NLO with and without hadronization corrections, 
and the predictions of the QCD model RAPGAP.  The shaded band shows 
the renormalization scale uncertainty of the NLO calculation. {\bf b} 
The ratios of the data and various predictions. The vertical error 
bars correspond to the errors of the data only.
\label{fig:h1-dijets-y2}}
\end{figure}

The normalized dijet cross section as a function of \yt~is compared in 
Fig.~\ref{fig:h1-dijets-y2} with an NLO QCD prediction~\cite{disent}.
The calculation uses the CTEQ5M parton density functions (pdf) and $Q$
as the renormalization and factorization scale.  The value of
\alpsmz~is set to 0.1183.  In addition the LO prediction, the NLO
prediction corrected for hadronization effects, and the dependence on
the renormalization scale ($Q/2 \leq \mu_{R} \leq 2Q$) are shown.

We observe that for $\yt \geq 0.001$ the data are well described by
QCD, while for decreasing \yt~QCD increasingly overestimates the data. 
In the region of very small jet separation the difference between the
LO and NLO prediction as well as the renormalization scale dependence
and the hadronization correction are all large.  This suggests that
fixed order perturbative QCD predictions are not reliable in the
region $\yt < 0.001$.  The deviation of NLO QCD in the highest \yt~bin
is due to large sensitivity to the pdf in this region of dijet phase
space and suggests that they need improvement there.

The data are also compared to RAPGAP~\cite{rapgap} which models QCD 
using LO matrix elements matched to parton showers and which uses the 
Lund string model for hadronization. It provides an excellent 
description of the data over the full \yt~range, in particular 
describing well the low \yt~region of multiple parton emissions.

\begin{figure}[htb]
\epsfxsize=14pc 
\epsfysize=25.5pc
\epsfbox{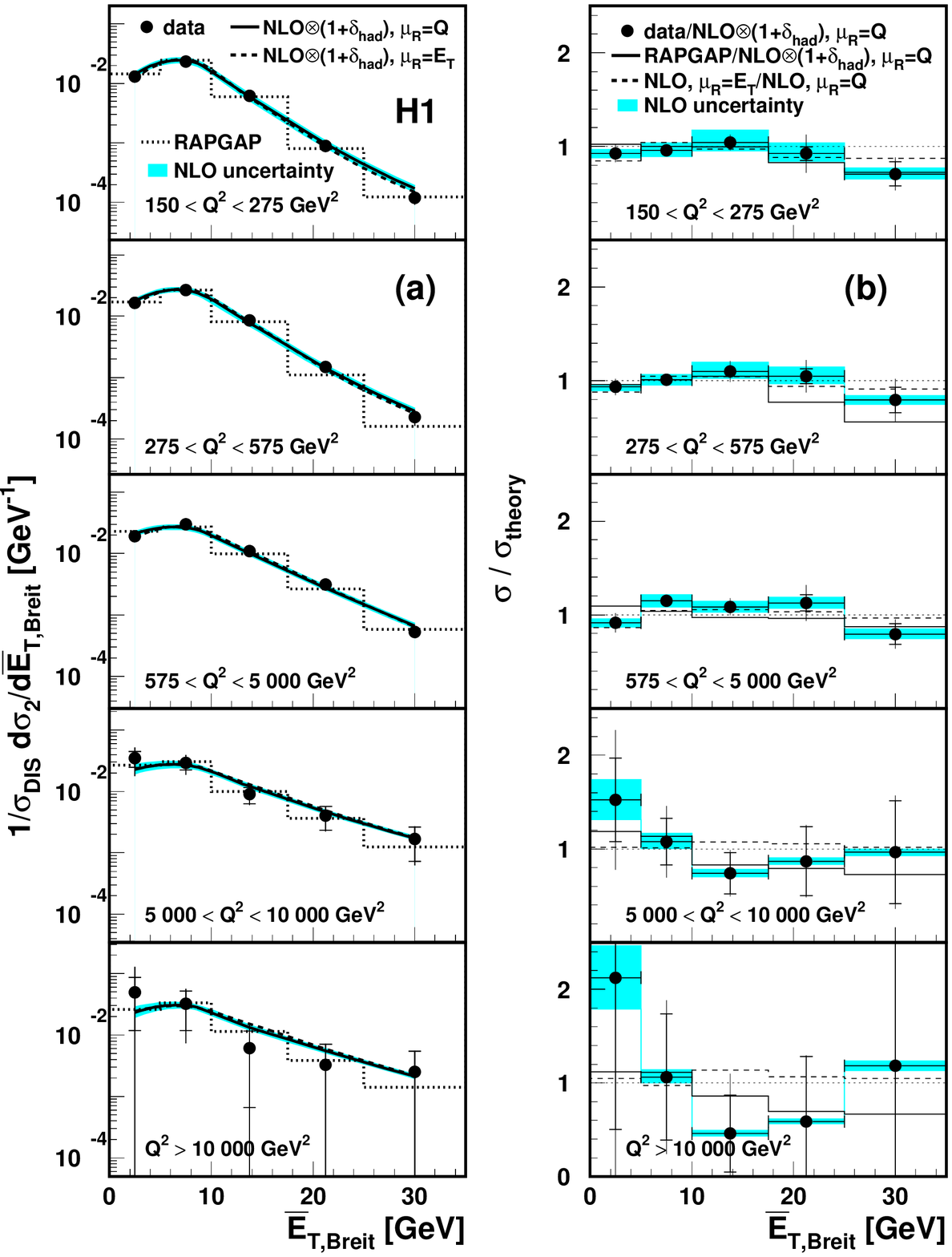}
\epsfxsize=7.7pc \epsfysize=25.5pc
\epsfbox{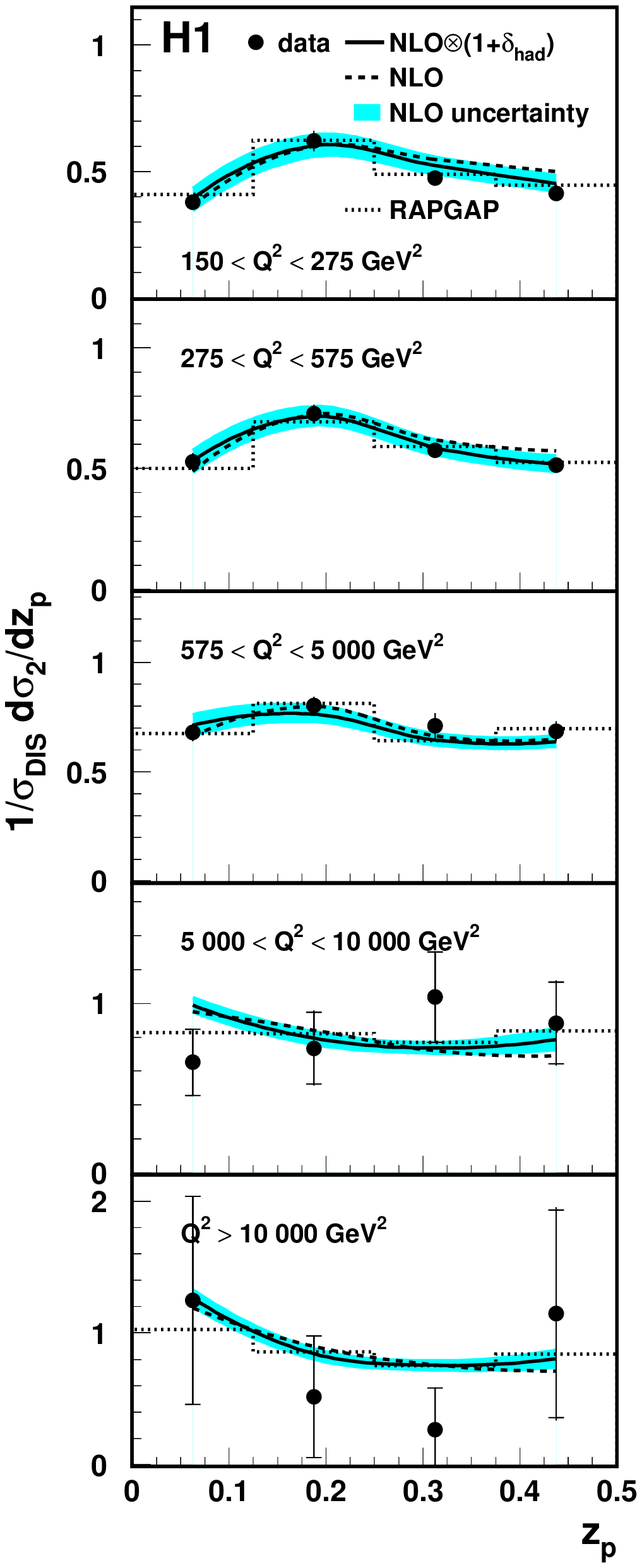} 
\epsfxsize=7.7pc \epsfysize=25.5pc
\epsfbox{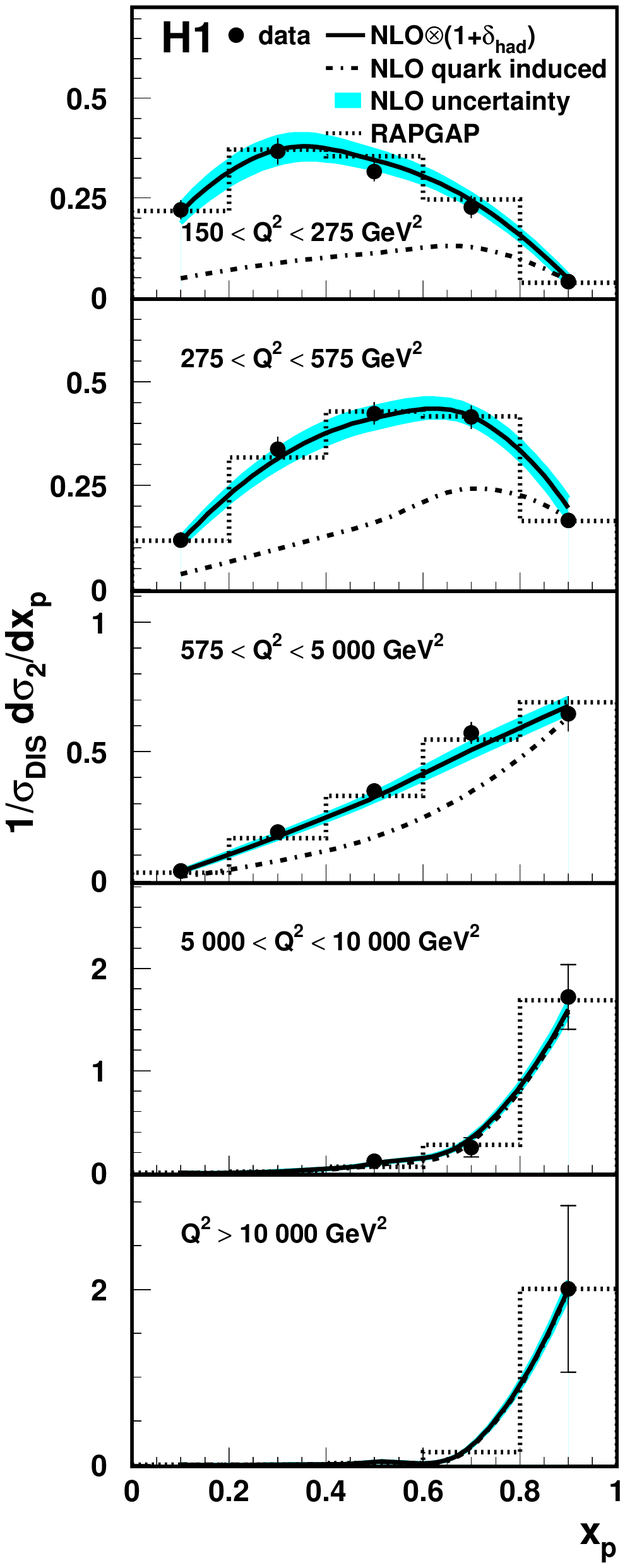} 
\caption{ The normalized dijet cross sections.  The predictions of NLO
QCD and RAPGAP are shown.  The shaded band corresponds to the
quadratic sum of the hadronization and renormalization scale
uncertainties.  {\bf a} Besides the NLO QCD predictions with the scale
$\mu_{R} = Q$, calculations with $\mu_{R} = \etbreit$ are also shown. 
{\bf b} The ratios of the data and various predictions. 
\label{fig:h1-dijets-etbreit-xp-zp}}
\end{figure}

Motivated by the agreement of NLO QCD with data for $\yt \geq 0.001$,
we investigate this sample of dijets, containing about 1/3 of the
selected DIS events, in more detail.  The following observables have
been investigated~\cite{h1-dijets}: the mean transverse energy of the
dijets in the Breit frame, the polar angle of the forward and backward
jet in the laboratory frame, and \xp~and \zp, which are frequently
used to express LO matrix elements.  They are calculated according to
$\xp = \frac{\QQ}{\QQ + m_{1,2}^{2}}$ and $\zp = \frac{\min_{i=1,2}
E_i(1-\cos\theta_i)}{\sum_{i=1,2}E_i(1-\cos\theta_i)}$.  In
Fig.~\ref{fig:h1-dijets-etbreit-xp-zp} the distributions in \etbreit,
\xp, and \zp~are shown.
The NLO QCD calculations describe these distributions well.  For
\etbreit it is shown that the differences between the two choices of
renormalization scale, $Q$ and \etbreit~are small.  There is a sizable
fraction of events with $\etbreit < 5$~GeV, which is however well
descibed by the predictions.  Also RAPGAP agrees well with the data. 
The quark-induced contribution of the NLO calculation is shown in
Fig.~\ref{fig:h1-dijets-etbreit-xp-zp} as a function of \xp.  It
varies from 30\% at the lowest \QQ~to almost 100\% in the highest
\QQ~bins.  This illustrates that these measurements are sensitive to
both the quark and the gluon-initiated contributions to the cross
section.

\section{Summary}

The minimum jet separation has been investigated for which NLO QCD can
give a good description of dijet production in DIS in the range $150 <
\QQ < 35000$~GeV$^{2}$ and $0.1 < y < 0.7$.  The required jet
separation is found to be small, \ie $\yt = 0.001$, selecting a dijet
sample containing 1/3 of the DIS events, significantly larger than the
approximately 1/10 obtained with more typical selection criteria.  The
good description obtained holds for either choice of renormalization
scale $Q$ or \etbreit~and covers regions in which both gluon and quark
induced processes dominate.

These measurements are also a significant challenge to QCD Monte Carlo 
models, which besides LO martix elements and parton showers also 
include hadronization and particle decays. RAPGAP is found to give a 
good description of all data.


\end{document}